\documentstyle[conf_iap,]{article}
\begin{document}
\heading{%
%
Quantum vacua:\\
Momentum space topology of fermion zero modes
%
}
\par\medskip\noindent
\author{%
G.E. Volovik$^{1,2}$
}
\address{%
Low Temperature Laboratory, Helsinki University of
Technology\\
P.O.Box 2200, FIN-02015 HUT, Finland\\
$^{2}$L.D. Landau Institute for
Theoretical Physics\\ Kosygin Str. 2, 117940 Moscow, Russia
}

\begin{abstract}
Quantum vacua are characterized by the topological structure of their 
fermion zero modes. The
vacua are distributed into universality classes protected by topology 
in momentum space. The
vacua whose manifold of fermion zero modes has co-dimension 3 are of 
special interest because in
the low-energy corner the fermionic excitations become the  Weyl relativistic
chiral fermions, while the dynamical bosonic collective modes of the 
fermionic vacuum
interact with the chiral fermions as the effective gravity and gauge fields.
The relativistic invariance, the chirality of fermions, the gauge and 
gravity fields, the
relativistic spin, etc., are the emergent low-energy properties of 
the quantum vacuum with such
fermion zero modes. The vacuum of the Standard Model and the vacuum 
of superfluid $^3$He-A belong
to this universality class and thus they are described by similar 
effective theories. This allows
us to use this quantum liquid for the theoretical and experimental 
simulations of many problems
related to the quantum vacuum, such as the chiral anomaly and the 
cosmological constant problems.
\end{abstract}
\section{Universality classes of fermionic vacua}
\subsection{Fermion zero modes of quantum vacuum}
All known elementary particles and gauge bosons are either massless 
or have the mass which is
extremely small compared to the Planck energy scale. This suggests 
that all the particles and
fields  originate from the fermionic or bosonic zero modes of the 
quantum vacuum. If so, our goal
must be to describe and classify the possible zero modes of quantum 
vacua and their interactions.
This does not require the knowledge of the exact `microscopic' 
(trans-Planckian) structure of the
quantum vacuum and can be done on the phenomenological grounds.

The classification of quantum vacua in terms of their zero modes is 
simplified if we assume that
only fermions are fundamental, while the bosons are composite 
representing the collective modes of
the fermionic quantum vacuum. As distinct from the symmetry 
classification of the vacuum states, it
is the topology of the propagator of the fermionic field which 
distinguishes between different
quantum vacua. The fermionic vacua and their fermion zero modes are 
thus distributed into
universality classes protected by momentum space topology. If the 
space dimension is D=3 and if we
consider only translationally invariant vacua, there are only two 
universality classes of
fermion zero modes which are protected by topology. The topologically 
stable zeros of the fermionic
energy spectrum form in the 3D momentum space the manifolds of either 
co-dimension 1 or
co-dimension 3. The universality class corresponding to co-dimension 
1 contains the vacua whose
fermion zero modes form the 2D surface in the 3D momentum space -- 
the Fermi surface. The class of
co-dimension 3 contains the vacua whose fermion zero modes are 
concentrated near the distinguished
topologically stable points in the 3D momentum space -- the Fermi points.

The elementary particles of our quantum vacuum originate from the 
massless chiral Weyl fermions
of the Standard Model, left and right quarks and lepton. The Weyl 
fermions are typical
representatives of the fermion zero modes of co-dimension 3. The 
Hamiltonian for the $a$-th chiral
fermionic species is
\begin{equation}
{\cal H}_a=cC_a \sigma^ip_i ~,
\label{neutrino}
\end{equation}
where $C_a=\pm 1$ is the chirality of the
fermion. This Hamiltonian vanishes at the point ${\bf p}=0$ in 3D 
momentum space.

\subsection{Momentum space topological invariant for Fermi point}

To characterize this singular point in the 3D momentum space (or any 
other manifold of zeros of
co-dimension 3) let us introduce a somewhat more
general
$2\times 2$ Hamiltonian which does not obey the Lorentz invariance:
\begin{equation}
{\cal H} =
   \sigma^i M_i({\bf p})~.
\label{MofP}
\end{equation}
Here ${\bf M}$ is an arbitrary function of the momentum ${\bf p}$.
The vector field ${\bf M}({\bf p})$ can have hedgehogs -- the points where
  $|{\bf M}({\bf p})|=0$ and thus the fermionic energy $E^2({\bf 
p})=|{\bf M}({\bf
p})|^2=0$. Such a point is topologically stable if the following 
integral over the 2-surface
$\sigma_2$ around this point is non-zero:
\begin{equation}
N_3= {1\over 8\pi}e_{ijk}\int_{\sigma_2} dS^k ~{1\over |{\bf M}({\bf 
p})|^3 } {\bf
M}\cdot \left({\partial  {\bf M}\over\partial {p_i}} \times {\partial
  {\bf M}\over\partial {p_j}}\right)~.
\label{NeutrinoInvariant}
\end{equation}
In the case of chiral fermions in Eq.~(\ref{neutrino}), the field 
${\bf M}({\bf p})=c{\bf p}$
represents the hedgehog in momentum space, and the topological
charge of this hedgehog coincides with the chirality of fermions, 
$N_3=C_a=\pm 1$. Invesigation
of the topological properties of the Fermi points was started in 1981 
for fermions on a lattice
\cite{NielsenNinomiya}, where these points were referred to as 
generic degeneracy points, and in
1982 in superfluid $^3$He-A \cite{VolovikMineev1982} where they were 
called boojums on Fermi
surface. According to Abrikosov and Beneslavskii
\cite{AbrikosovBeneslavskii1971} and Nielsen and Ninomiya 
\cite{NielsenNinomiya2} such points can
also appear in semiconductors as the crossing (diabolic) points of 
two energy bands. In condensed
matter examples the spin is effective: the Pauli matrices $\sigma^i$ 
act in the space of two
crossing bands.

\section{Emergent relativistic quantum field theory}

\subsection{Emergence of chiral fermions, gauge field and gravity}

The Fermi points with the elementary topological charge, $N_3=\pm 1$, 
have a remarkable property.
Near such a point $p_i^{(a)}$ the vector field ${\bf M}({\bf p})$ and 
the Hamiltonian have the
following general form:
\begin{equation}
   M_n({\bf p})\approx e^i_n(p_i-p_i^{(a)})~~,~~ {\cal H}_a =
   e^i_n \sigma^n (p_i-p_i^{(a)})~.
\label{MofPexpansion}
\end{equation}
After the shift of the momentum and the diagonalization of the 
$3\times 3$ matrix
$e^i_n$, one again obtains ${\cal H}_a=C_ac \sigma^ip_i$, where the
chirality $C_a$ is determined by the sign of the determinant of the 
matrix $e^i_n$. This means that
fermions near such a Fermi point are always the  Weyl fermions -- the 
relativistic chiral
fermions -- even if the original system is not relativistic. The 
quantities $e^i_n$ (or
the effective metric $g^{ik}=\sum_n e^i_ne^k_n$) and
$p_i^{(a)}$ emerging at low energy are dynamical variables 
characterizing the bosonic collective
modes of the fermionic vacuum. They interact with chiral fermions as 
the effective gravity and
gauge field, respectively.

Thus the relativistic invariance, the chirality of fermions, the 
gauge and gravity fields, and also
the relativistic spin are the emergent low-energy properties of the 
quantum vacuum with fermion
zero mode of co-dimension 3, if its topological charge is elementary,
$N_3=+1$ or $N_3=-1$.

\subsection{Role of discrete symmetries: \\emergence of parity and 
non-Abelian local symmetry}

What happens if the topological charge of the Fermi point is not 
elementary? For instance,
the positions of two Fermi points with the charges  $N_3=-1$ and 
$N_3=+1$ can coincide, so that the
total topological charge of the singular point is trivial, $N_3=0$. 
It appears that all the above
properties of the emerging relativistic quantum field theory will 
survive if there is some discrete
symmetry between these two fermionic species which protects zeros of 
the Hamiltonian. In the
low-energy `relativistic' limit this discrete symmetry manifests 
itself as the space parity ${\rm
P}$ which transforms the left-handed fermion to the right-handed one. 
If this discrete symmetry is
violated or spontaneously broken, the zeros in momentum space are 
protected neither by topology nor
by symmetry; zeros disappear which means that in the relativistic 
low-energy limit the
fermion zero modes acquire the Dirac mass. Such situation occurs in 
the Standard Model and
in the so-called planar phase of superfluid $^3$He. In the planar 
phase, the corresponding space
parity ${\rm P}$ of the low-energy fermions is effective: it evolves 
from some approximate internal
symmetry of the `high-energy' atomic physics of the quantum liquid.

Another important role of discrete symmetries shows up if we consider the
Fermi point with the multiple charge, say $N_3=+2$. If there is a 
proper discrete
symmetry between the fermions, the singular point becomes the 
combination of two
elementary  Fermi points describing two right-handed fermionic 
species coupled by this symmetry. In
this case the effective relativistic invariance is not distorted, but 
in addition one finds that
the double degenerate Fermi point generates the effective non-Abelian 
gauge field. Since the
positions
${\bf p}^{(1)}$ and
${\bf p}^{(2)}$ of constituent Fermi points with $N_3=+1$ can 
oscillate separately, the $4\times 4$
Hamiltonian for two fermionic species becomes
\begin{equation}
  {\cal H} = e^i_n \sigma^n (p_i-A_i -\tau_b A^b_i)~.
\label{4times4}
\end{equation}
Here $\tau_b$ is the Pauli matrix in the isotopic space of two 
species of fermion zero modes, and
the new collective mode $A^b_i$ plays the role of the Yang-Mills 
field. Such situation occurs in
the Standard Model and in the superfluid $^3$He-A. The higher 
discrete symmetry groups of the Fermi
points can generate the higher-order effective local symmetry in the 
fermionic sector. For example
the Fermi point with $N_3=+4$ and with the discrete symmetry
$Z_4$ or $Z_2\times Z_2$ will lead to the local $SU(4)$ symmetry in 
the low-energy corner.

According to Eq.~(\ref{4times4}), in the vicinity of Fermi points the 
fermionic sector acquires the
local $U(1)$ and non-Abelian symmetry groups, and also the general 
covariance. This, however, does
not imply that such symmetries will automatically emerge in the 
bosonic sector, i.e. in the
effective action for the collective fields $A_i$, $A^b_i$ and
$e^i_n$. But this is possible at least in
principle, since the corresponding action for the bosonic fields come 
from the integration over the
fermionic fields in the same manner as in Sakharov's induced gravity 
\cite{Sakharov}. The problem
is that because of the ultraviolet divergences the bosonic action is 
generated not only by
fermion zero modes but also by the high-energy degrees of freedom of 
the fermionic vacuum which are
not necessarily Lorentz invariant (see the detailed discussion in Ref.
\cite{VolovikBook}).

\subsection{Chiral anomaly}

There are some properties of the quantum vacuum which do not depend 
on whether the bosonic sector
of the effective theory has all the symmetries of the fermionic 
sector or not. An example is
provided by the chiral anomaly, which is completely determined by the 
spectral flow through the
Fermi point and thus by its momentum space topology. The fermionic 
charge $B$ produced per unit
time per unit volume from the vacuum in the presence of the effective 
magnetic and electric fields
is given by the following generalization of the equation derived by Adler
\cite{Adler} and Bell and Jackiw \cite{BellJackiw} for axial anomaly:
\begin{equation}
\dot B ={1\over {16\pi^2}}
F_{\mu\nu}F^{*\mu\nu} \sum_a N_{3a} B_a q_a^2~.
\label{ChargeParticlProduction}
\end{equation}
Here $q_a$ is the charge of the
$a$-th fermion with respect to the effective gauge field $F^{\mu\nu}$;
$F^{*\mu\nu}=(1/2)e^{\alpha\beta\mu\nu}F_{\alpha\beta}$ is the dual
field strength; and $N_{3a}$ is the topological charge of the Fermi point. This
relativistic, gauge invariant and covariant equation is applicable to 
all systems with Fermi
points including $^3$He-A where the bosonic sector certainly does not 
obey these symmetries.
In $^3$He-A the  Adler--Bell--Jackiw equation has been  verified in 
experiments with continuous
vortex-skyrmions
\cite{BevanNature}.

\section{Vacuum energy}

The fermion zero modes of quantum vacuum can say nothing on the 
microscopic properties of the
quantum vacuum, such as the value of the vacuum energy and thus of 
the cosmological constant
$\rho_\Lambda$. The naive estimation of the vacuum energy density, as 
the zero-point energy
  $(1/2)E(p)$ of bosonic
modes plus the negative energy of fermion modes in
the Dirac sea, gives
\begin{equation}
\rho_\Lambda \sqrt{-g}= {1\over V}\left(\nu_{\rm bosons}\sum_{\bf p}{1\over 2}
cp~~ -\nu_{\rm fermions}
\sum_{\bf p} cp\right)~,
\label{VacuumEnergyPlanck}
\end{equation}
where $V$ is the volume of the system; $\nu_{\rm bosons}$ and and 
$\nu_{\rm fermions}$ is the
number of bosonic and fermionic zero modes. Because of the 
ultraviolet divergence this estimate is
in a huge disagreement with observations which is known as the 
cosmological constant problem.

To calculate the real energy of the vacuum state one must know the 
ultraviolet (microscopic)
structure of the given vacuum. We have no information on our quantum 
vacuum, but we can consider
as a guide the well known quantum vacua of the same universality 
class. The result is rather
unexpected and so general that it can be applicable to any vacuum.
It appears that the contribution of zero modes in 
Eq.~(\ref{VacuumEnergyPlanck}) is
completely cancelled by the contribution from all other microscopic 
degrees of freedom without any
fine tuning.  Applying this to our vacuum one may conclude that the 
equilibrium quantum vacuum
does not gravitate.

This cancellation is the result of the local stability of the vacuum 
state, and it is exact if the
vacuum is not perturbed. The vacuum responds to perturbations, and 
the energy density of the
perturbed vacuum is on the order of the energy density of 
perturbations. Thus the cosmological
constant is not a constant at all but is the dynamical quantity which
is either continuously or in a stepwise manner adjusted to perturbations.
Since all the perturbations of the vacuum in the present Universe 
(gravitating matter, expansion,
curvature, non-zero temperature, inhomegeneity, etc.) have energy 
scales much lower than the Planck
scale, our vacuum is extremely close to the local equilibrium, and 
thus the dark energy today is
extremely small.

What happens if a phase transition occurs in which the symmetry
of the vacuum is broken, as is supposed to happen in the early Universe
when, say, the electroweak symmetry was broken? In the effective
theory, such a transition must be accompanied by a change of the vacuum
energy, which means that the vacuum must have a huge energy either 
above or below the phase
transition. However, the exact microscopic theory suggests the
phase transition does not disturb the zero value of the vacuum energy.
After the vacuum relaxes to a new equilibrium state,
its energy density will be zero again. The energy change is 
completely compensated by the
change of the internal (microscopic) quantities characterizing the 
microscopic structure of the
vacuum. In the quantum vacuum of quantum liquids this is the chemical 
potential and density of the
atoms of the liquid \cite{VolovikBook}.

\section{Other systems}

We considered here the effective theory of fermion zero modes 
emerging in the low-energy corner
of the quantum vacuum belonging to the universality class of the 
co-dimension 3. This class is
of special interest because the corresponding effective theory is the 
quantum field theory of
chiral fermions interacting with gauge fields and gravity. However, 
in the condensed matter physics
the more popular is the universality class of the co-dimension 1. It 
contains the Fermi liquids
with Fermi surfaces. The effective theory of such vacua has been 
constructed by Landau \cite
{Landau2} and is known now as the Landau theory of Fermi liquids. A 
similar effective theory
emerges for fermion zero modes in the core of quantized vortices; in 
the mixed state of
superconductors; and in relativistic theories in the presence of 
strong fields, for instance, in
the vicinity and beyond the black-hole horizon. In multi-dimensional 
systems the higher
universality classes are also topologically protected, such as the 
class of co-dimension 5; in
addition the fermion zero modes of the co-dimension 3 appear on 
branes, and all the properties of
the relativstic quantum field theory emerge for the brane matter at 
low energy. In the systems
with even space dimension the fully gapped vacua also have 
non-trivial momentum space topology,
which leads to quantization of physical parameters, and other exotic 
phenomena. Discussion of
all these universality classes of quantum vacua can be found in Ref.
\cite{VolovikBook}.

\vfill
\end{document}